\documentclass[conference]{IEEEtran}
\usepackage{cite}
\usepackage{amsmath,amssymb,amsfonts}
\usepackage{algorithmic}
\usepackage{multirow}
\usepackage{graphicx}
\usepackage{textcomp}
\usepackage{xcolor}
\def\BibTeX{{\rm B\kern-.05em{\sc i\kern-.025em b}\kern-.08em
    T\kern-.1667em\lower.7ex\hbox{E}\kern-.125emX}}
\begin{document}

\title{Reconstructing Speech from Real-Time Articulatory MRI Using Neural Vocoders
\\
}

\author{ \IEEEauthorblockN{Yide Yu}
\IEEEauthorblockA{\textit{Institute of Informatics} \\
\textit{University of Szeged}\\
Szeged, Hungary \\
mr\_yideyu@163.com} \and
\IEEEauthorblockN{Amin Honarmandi Shandiz}
\IEEEauthorblockA{\textit{Institute of Informatics} \\
\textit{University of Szeged}\\
Szeged, Hungary \\
shandiz@inf.u-szeged.hu} \and
\and \IEEEauthorblockN{L\'aszl\'o T\'oth}
\IEEEauthorblockA{\textit{Institute of Informatics} \\
\textit{University of Szeged}\\
Szeged, Hungary \\
tothl@inf.u-szeged.hu} 
}

\maketitle

\begin{abstract}
Several approaches exist for the recording of articulatory movements, such as eletromagnetic and permanent magnetic articulagraphy, ultrasound tongue imaging and surface electromyography. Although magnetic resonance imaging (MRI) is more costly than the above approaches, the recent developments in this area now allow the recording of real-time MRI videos of the articulators with an acceptable resolution. Here, we experiment with the reconstruction of the speech signal from a real-time MRI recording using deep neural networks. Instead of estimating speech directly, our networks are trained to output a spectral vector, from which we reconstruct the speech signal using the WaveGlow neural vocoder. We compare the performance of three deep neural architectures for the estimation task, combining convolutional (CNN) and recurrence-based (LSTM) neural layers. Besides the mean absolute error (MAE) of our networks, we also evaluate our models by comparing the speech signals obtained using several objective speech quality metrics like the mean cepstral distortion (MCD), Short-Time Objective Intelligibility (STOI), Perceptual Evaluation of Speech Quality (PESQ) and Signal-to-Distortion Ratio (SDR). The results indicate that our approach can successfully reconstruct the gross spectral shape, but more improvements are needed to reproduce the fine spectral details.
\end{abstract}


\section{Introduction}

Human speech production is a complex process that requires precisely coordinated movements from the respiratory organs, the larynx and the articulators. Theoretically, the configuration of the articulatory organs determines the speech signal that is being produced. Articulatory-to-acoustic mapping seeks to determine whether it is possible to estimate the speech signal if we know the physical positions of these organs. During the last decade, several types of devices have been proposed to record the movement of the articulators. The most important of these are 
ultrasound tongue imaging
(UTI)~\cite{Hueber2011,Jaumard-Hakoun2016,csapo2017c,Xu2017,Tatulli2017, Saha-ultra2speech},
electromagnetic articulography
(EMA)~\cite{Wang2012a,Wang2014,Kim2017a,Taguchi2018articulatory},
permanent magnetic articulography
(PMA)~\cite{Fagan2008,Gonzalez2016}, and surface electromyography
(sEMG)~\cite{Nakamura2011,Deng2014,Diener2015,Meltzner2017,Wand2018domainadversarial}.
While articulatory-to-acoustic mapping can also be considered as a theoretical problem, it has an important application, which is the creation of silent speech interfaces~\cite{Denby2010}. 
Silent speech interfaces (SSI) seek to generate speech from the articulatory movement without actually having access to the speech signal, with the aim of aiding the communication of speaking impaired people who can move their articulators, but cannot actually produce speech. SSIs may also be applied in situations where the speech signal cannot be heard, as in extremely noisy industrial environments and certain military applications.

Currently, the devices listed above typically record noisy and low resolution signals. Magnetic resonance imaging (MRI) may offer an alternative solution, as it can produce high resolution images. Moreover, while the other methods capture only the lingual, labial and jaw motions, MRI also covers the pharyngeal and nasal regions, which are not reachable by the other methods~\cite{Ramanarayanan2018}. Of course, MRI is costly and it requires a huge machine, so currently this technology cannot provide a basis for wearable silent speech interfaces, but treating the articulatory-to-acoustic mapping problem with MRI as input is a very interesting theoretical research topic in itself. The current developments of MRI technology allow us to create real-time recordings of the speech production process at rates of about 20-50 frames per second. While this is lower than the sampling rate of other tools such as ultrasound tongue imaging, the spatial resolution of MRI is typically much better~\cite{Narayanan2019}. Therefore, compared to the other tools, it provides complementary information, and thus examining the applicability of MRI recordings for articulatory-to-acoustic conversion may lead to some intriguing conclusions.

Several previous studies have applied articulatory MRI for speech-related tasks. The majority of these tried to perform speech recognition using the MRI as input~\cite{douros2018rtmri, Katsamanis2011, Saha2018mri}. Interestingly, some authors tried to estimate the MRI from the speech signal, which is the inverse of the problem we study here~\cite{LiMRI}. However, relatively few attempts have been made to synthesize speech signals based on the MRI~\cite{Toutios2016, csapotMRI}. Here, we extend the recent study by Tamás G. Csapó~\cite{csapotMRI}. Similar to his approach, we use the MRI recordings of the USC-TIMIT data set as input~\cite{USC-TIMIT}. However, while he used a simple conventional vocoder for the speech synthesis process, here we apply more recent neural vocoders for this task. As the name suggests, these are based on deep learning models, and they have been reported to produce higher quality speech~\cite{govalkar}. Hence, the we approach we present here is purely neural. In the next section we shall introduce the concept of our articulatory-to-acoustic mapping framework, with a special focus on the application of neural vocoders.

\section{MRI-based Articulatory-to-Acoustic Conversion}

\subsection{Real-Time Articulatory MRI}
\label{sec:MRI}

As the input MRI data we used the freely available USC-TIMIT data set~\cite{USC-TIMIT}. This data set contains synchronized speech and real-time MRI recordings from American English speakers. The subjects lying in an MRI device were asked to read 460 sentences from the MOCHA-TIMIT database. The MRI data was captured using an 1.5 Tesla Sigma Excite HD MRI scanner. The resolution of the recorded images is 64x64 pixels, and their orientation is adjusted to the mid-saggital plane. The time resolution of the recordings was 23 frames per second. The speech signal was recorded simultaneously with the MRI video inside the MRI scanner at a sampling rate of 20 kHz. Due to the noisy operation of the scanner, the recorded speech signals were very noisy, so they had to be post-processed using noise cancellation algorithms.

\begin{figure}
\centering
\includegraphics[width=0.4\textwidth]{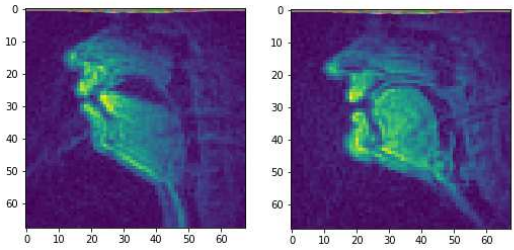}
\caption{\textit{Sample MRI images from two speakers.}} \label{fig:MRI_sample}
\end{figure}

Fig~\ref{fig:MRI_sample} shows two example images from a male and a female speaker from the database. As can be seen, the images have significant inter-speaker differences, so in the experiments we created speaker-dependent models, that is, separate models for each speaker. The database contains samples from five male and five female subjects, and we worked with the data of two males and two females. Since Csapó reported earlier that the data of speaker 'M1' has significant misalignment problems~\cite{csapotMRI}, we worked with speakers 'M2', 'M3', 'F2' and 'F3'. 

\subsection{Speech Generation using Neural Vocoders}
\label{sec:vocoders}
\begin{figure}
\centering
\includegraphics[width=0.3\textwidth]{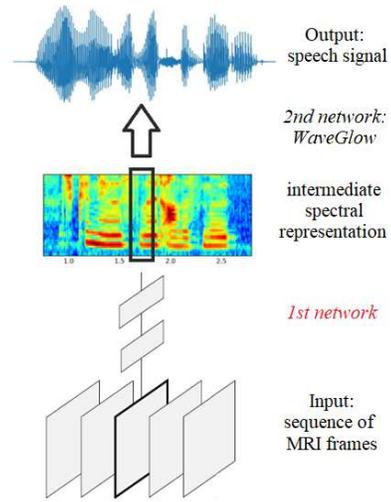}
\caption{\textit{Schematic diagram of the MRI-to-speech conversion process applied here. Out goal is to find the optimal structure and parameters for the first network of the processing chain.}} \label{fig:framework}
\end{figure}

As shown in Fig~\ref{fig:framework}, our goal is to convert a sequence of MRI images into a speech signal. This task can be formalized as a sequence-to-sequence mapping problem. For such tasks, various sophisticated deep neural network (DNN) structures have been proposed recently that do not even require aligned data. However, our MRI and speech sequences were synchronized, allowing us to use much simpler architectures that estimate the mapping in a pairwise manner, that is, give one output vector for each element of the input sequence.

\begin{table*}[t]
\caption{The layers of the 3 network architectures, along with their main parameters. To save space, some details such as dropout layers and regularization parameters are not shown.}\label{tab:ann}
\centering
\renewcommand{\arraystretch}{1.1} 
\begin{tabular}{|l|l|l|}
\hline
\hfil2D-CNN+BiLSTM & \hfil3D-CNN & \hfil3D-CNN+BiLSTM \\
\hline
TimeDistributed(Conv2D(30,(13,13),strides=(2,2))&Conv3D(30, (5, 13, 13), strides=(sts, 2, 2) & Conv3D(30, (5, 13, 13), strides=(sts, 2, 2) \\
TimeDistributed(Conv2D(60,(13,13),strides=(2,2))&Conv3D(60, (1, 13, 13), strides=(1, 2, 2) & Conv3D(60, (1, 13, 13), strides=(1, 2, 2) \\
TimeDistributed(MaxPooling2D((2,2)))&MaxPooling3D((1, 2, 2)) & MaxPooling3D((1, 2, 2)) \\
TimeDistributed(Conv2D(90,(13,13),strides=(1, 1))&Conv3D(90, (1, 13, 13), strides=(1, 1, 1)&Conv3D(90, (1, 13, 13), strides=(1, 1, 1) \\
TimeDistributed(Conv2D(85,(13,13),strides=(2,2))&Conv3D(120, (5, 3, 3), strides=(1, 2, 2) &Conv3D(120, (5, 3, 3), strides=(1, 2, 2) \\
TimeDistributed(MaxPooling2D((2,2))&MaxPooling3D((1, 2, 2)) &MaxPooling3D((1, 2, 2)) \\
TimeDistributed(Flatten())&Flatten() &Reshape((5, 480)) \\
Bidirectional(LSTM(320)&Dense(1000)&Bidirectional(LSTM(370) \\
Dense(80, activation='linear')&Dense(80, activation='linear')&Dense(80, activation='linear') \\
\hline
\end{tabular}
\label{table_nets}
\end{table*}

Thus, the input of our network is an MRI image, or a short sequence of consecutive images. However, the optimal choice for the output or target vector is not trivial. Creating DNNs that output speech signals directly is feasible~\cite{melgan, waveglow}. However, such networks generate tens of thousands of speech samples per second, and, consequently, these models are enormous and their training requires a huge amount of speech data. This is why we applied an alternative, indirect approach here, which is depicted in Fig~\ref{fig:framework}. This approach was motivated by neural speech synthesis, where DNNs are applied in two steps~\cite{waveglow}. First, there is a network that estimates a spectral representation from the text to be synthesized. Then, there is a second network that generates the speech signal from the spectral representation. Adapting this approach to our task requires modifications only in the first network, as our input is an MRI video and not a text. However, the second step is the same, so we can borrow large, pre-trained networks for the second task from text-to-speech synthesis. Doing so, we had to create and train only the first network, which had to estimate a dense spectral representation instead of the speech waveform itself. As for the second, the speech generation task, several neural vocoders are available~\cite{govalkar}, and we chose to use the WaveGlow model~\cite{waveglow}, as it worked well in our earlier study~\cite{Csapo2020}. WaveGlow requires a sequence of 80-dimensional mel-scaled spectral vectors as the input, so our task was to create a network that can estimate such a spectral vector for each frame of the MRI video. As the default frame rate of WaveGlow (86 fps) is almost 4 times higher than the frame rate of our MRI video (23 fps), we estimated the missing spectral vectors by applying interpolation before the actual speech synthesis step.

\section{DNN Architectures for MRI-to-Spectrogram Conversion}

To prepare the MRI images for DNN training, the pixel intensities of each image were min-max normalized to the $[-1,1]$ range, and the speech recordings were resampled at 22050 Hz, as this is the sampling rate required by WaveGlow. The speech signals were then converted to a mel-spectral representation, and the resulting 80-dimensional mel-spectral vectors were standardized before using them as the DNN training targets. From the 92 sentences of each subject, 4 were used for validation and 2 for testing.

Formally, our networks has to map each MRI image to a spectral vector. However, using several consecutive input frames instead of a single frame can significantly improve the results~\cite{toth20203d, Saha-ultra2speech}. Hence, the input for all our network configurations was a 3D array, treating time as the the third axis besides the two spacial axes of the images. Table~\ref{tab:ann} summarizes the main parameters of our experimental network configurations, and we also give a brief description of each below.  Our networks were implemented in Keras. We applied the Swish activation
function~\cite{swish} in the hidden layers. The input window of the network contained 13 consecutive video frames. The number of trainable parameters was approximately the same for all 3 configurations. For the loss function we applied the mean absolute error (MAE), as it was reported to give slightly better results than the more conventional mean-squared error for speech-related tasks~\cite{pandeyl1}.

{\bf 2D-CNN+BiLSTM:} When working with a sequence of images, a popular technique is to process each image using a convolutional neural network (CNN), and then combine the results along the time axis using recurrent neural structures such as the long short-term memory (LSTM) layer~\cite{hochreiter1997lstm}. Thus, our first network combined 2D-CNN layers that process each image with an LSTM layer on top to fuse the information along the time axis. As shown in Table~\ref{tab:ann}, the 2D convolutional and max-pooling layers are applied to each image of the input sequence using the TimeDistributed() function of Keras. Then their outputs are combined along the time axis using an LSTM layer, and the actual regression task with respect to the spectral target vectors is performed by the topmost linear layer.

{\bf 3D-CNN:} Several authors argued recently that good video classification is also achievable using purely convolutional structures by extending the convolution to the temporal axis~\cite{Tran,Saha-ultra2speech,toth20203d}. This is why a 3D convolutional model served as our second model (see Table~\ref{tab:ann}). Its lowest layer processes the input in 5-frame blocks with a hop size controlled by the $sts$ parameter. This parameter allows us to analyze input blocks that are placed at bigger time intervals. In an earlier study we processed ultrasound videos which had a much larger frame rate, and the optimal value for $sts$ was found to be 5~\cite{toth20203d}. Here, we got the best performance with $sts=3$, which is reasonable as the frame rate was much lower. The subsequent Conv3D layers essentially process each input block separately (their filter size along the time axis being set to 1), and the uppermost Conv3D layer performs the fusion along time. This structure was motivated by the findings of Tran et al.~\cite{Tran}. Finally, the regression is performed by a Dense layer.

{\bf 3D-CNN+BiLSTM:} A drawback of the 2D-CNN+BiLSTM model is that it cannot skip input frames, while the drawback with the 3D-CNN model is that recurrent layers may be more effective in fusing information obtained at several points along the time axis. Hence, we also experimented with a third model that combines the advantages of the two previous architectures. As Table~\ref{tab:ann} shows, this model retained the basic architecture of the 3D-CNN network, but we replaced the uppermost Dense layer with an LSTM layer.

\section{Results and Discussion}

\begin{table}[b]
\caption{Mean absolute error values for the validation and test sets.} \label{tab:MAE}
\vspace{-2mm}
\centering
\renewcommand{\arraystretch}{1.1} 
\begin{tabular}{l||c|c|c}
        & \multicolumn{3}{c}{{Mean absolute error (validation / test)}} \\
\cline{2-4}
speaker & 2D-CNN+BiLSTM & ~~3D-CNN~~ &  3D-CNN+BiLSTM \\
\hline\hline
`F2' & 0.26 / 0.26 & 0.28 / 0.28 & 0.26 / 0.26 \\
`F3' & 0.48 / 0.40 & 0.48 / 0.40 & 0.45 / 0.40 \\
`M2' & 0.37 / 0.33 & 0.36 / 0.33 &  0.36 / 0.32 \\
`M3' & 0.30 / 0.32 & 0.31 / 0.32 & 0.29 / 0.32 \\
\hline
avg. & 0.35 / 0.33 & 0.36 / 0.33 &  0.34 / 0.33 \\

\end{tabular}

\vspace{4mm}
\caption{MCD scores for the test set.} \label{tab:MCD}
\vspace{-2mm}
\centering
\begin{tabular}{l||c|c|c}
        & \multicolumn{3}{c}{{Mel-Cepstral Distortion (dB)}} \\
\cline{2-4}
speaker & 2D-CNN+BiLSTM & ~~3D-CNN~~ & 3D-CNN+BiLSTM \\
\hline\hline
`F2' & 5.87 & 6.02 & 5.84 \\
`F3' & 5.75 & 5.74 & 5.84 \\
`M2' & 5.54 & 5.47 & 5.40 \\
`M3' & 5.06 & 5.09 & 4.95 \\
\hline
average & 5.56 & 5.58 & 5.51 \\

\end{tabular}

\end{table} 

\begin{table}[b]
\caption{Objective speech quality scores for the test set.} \label{tab:objective}
\vspace{-2mm}
\centering
\renewcommand{\arraystretch}{1.1} 
\begin{tabular}{l||c|c|c|c|}
speaker & STOI & PESQ & SDR \\
\hline\hline
'F2' & 0.51 & 1.67 & -22.7 \\
'F3' & 0.27 & 1.86 & -25.7\\
'M2' & 0.52 & 1.65 & -18.9 \\
'M3' & 0.41 & 1.81 & -22.4 \\
\hline
average & 0.43 & 1.75 & -22.4 \\

\end{tabular}

\end{table} 

\begin{figure*}[t]
\centering
\includegraphics[width=0.8\textwidth]{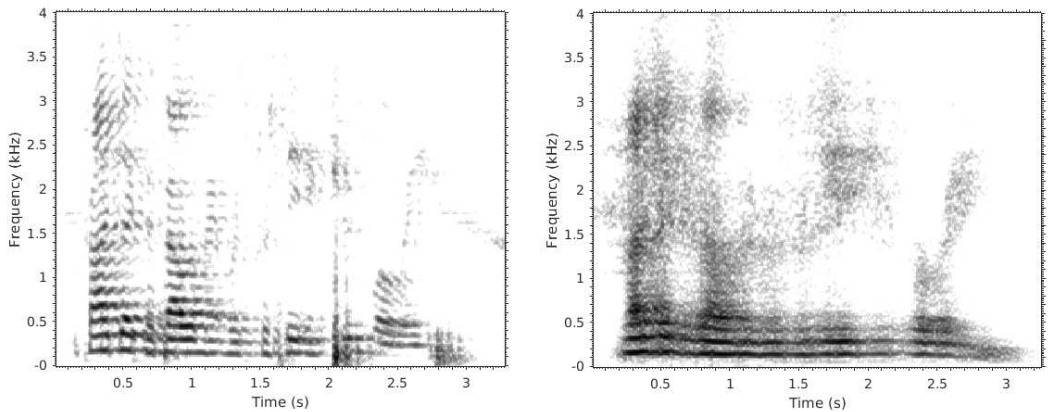}
\caption{\textit{Spectrograms of a sample sentence from speaker M3. Left: original, right: reconstructed from MRI.}} \label{fig:spect}
\end{figure*}

Table~\ref{tab:MAE} shows the MAE values obtained by the three network configurations on the validation and test sets of each speaker. The results indicate that there are large interpersonal differences compared to the average shown in the bottom row. However, the three networks architectures produced very similar results. Based on the average score on the validation set, the 3D-CNN+BiLSTM model seems to be slightly better than the 2D-CNN+BiLSTM network, and the latter is slightly better than the 3D-CNN, but the differences are negligible, and the average loss values are equivalent on the test set. 

To compare not only the DNN loss values but also the synthesized speech signals, we calculated the mel-cepstral distortion (MCD) of the test files. MCD is frequently used as an objective measure of speech quality in speech synthesis~\cite{MCD}. The MCD values reported in Table~\ref{tab:MCD} are not in complete accord with the MAE loss values obtained for the various speakers. This reflects the known fact that the simple DNN loss functions such as the MAE or MSE do not necessarily coincide with human perception~\cite{kolbaek2020loss}. However, the average MCD values have the same tendency as the MAE values in the sense that the 3D-CNN+BiLSTM model seems to be slightly better than the other two networks, but the difference is minimal.

To compare our results with those of the literature, Csapó performed a similar experiment using a conventional MGLSA vocoder, and he reported MCD scores around 4.5 for speakers 'F2' and 'M2'~\cite{csapotMRI}. However, he attempted to estimate only the spectral envelope, and he used the residual component of the original speech signal during the synthesis. In contrast, our approach attempts to reconstruct the full spectrum from the MRI only, including the fine spectral details such as the pitch information. To illustrate the performance of our solution in this respect,
we give an example of the narrow-band spectrogram (created from the output of the 3D-CNN+BiLSTM net) for speaker 'M3' in Fig.~\ref{fig:spect}. Comparing the estimated signal with the original, we see that our network is quite successful in reconstructing the rough spectral shape, but it fails with the fine spectral details. The horizontal stripes that reflect the fundamental frequency and its harmonics are preserved only in the lower spectral region. This is probably due the the mel-scale used by WaveGlow, as it represents the higher frequencies with a lower resolution. There is considerable smearing present along the time axis as well. For example. the plosive burst at 2 seconds in the original spectrogram is missing in the reconstructed signal. A further analysis is required to see whether the relatively low frame rate of the MRI is responsible for this.

We evaluated additional objective metrics to assess the quality and intelligibility of the synthesized signals (see Table~\ref{tab:objective}). The Short-Time Objective Intelligibility (STOI) metric~\cite{STOI} returns values between -1 and 1, so the average score of 0.43 we attained is fair, but not that good. The Perceptual Evaluation of Speech Quality (PESQ) metric~\cite{Martin-Donas} returns values on the mean opinion scale between 1 and 5. On this scale, our average score of 1.75 lies between "bad" and "poor". Lastly, the Signal-to-Distortion Ratio (SDR) metric~\cite{SDR} behaves much like the signal-to-noise ratio, so the negative values we got indicate that there is a large rate of distortion in our signals. However, we emphasize that the original speech recordings were also of low quality, as they were recorded in an MRI device, and they were post-processed by noise cancellation methods. 

\section{Conclusions}

Here, we investigated the feasibility of performing articulatory-to-acoustic mapping using neural vocoders such as WaveGlow. Compared to earlier attempts, the approach we proposed estimates the whole spectral content of the signal and not just the spectral envelope. Our experiments showed that our method is able to reconstruct the gross spectral shape and also some fine details such as the lowest pitch harmonics, but the spectrograms we got were quite blurred, and hence the resulting speech signals are at the lower end of the scale, according to several objective speech quality metrics. In the future we plan to incorporate perceptually motivated loss functions in the training process, as this would allow the direct optimization of the output speech quality~\cite{kolbaek2020loss}.


\section*{Acknowledgments}

This study was supported by grant NKFIH-1279-2/2020 of the Ministry for Innovation and Technology, Hungary, and by the Ministry of Innovation and the National Research, Development and Innovation Office by grant FK 124584 and the framework of the Artificial Intelligence National Laboratory Programme.  The GPU card used for the computations was donated by the NVIDIA Corporation. The authors are grateful for the comments and source code of Tamás Gábor Csapó.

%

\bibliographystyle{IEEEbib}
\bibliography{IJCNN2021}


\end{document}